\documentclass{article}

\usepackage{authblk}
\usepackage{abstract}

\usepackage{lscape}

\usepackage{graphicx}
\usepackage{mathptmx}

\usepackage{color}
\usepackage{caption}
\usepackage{subcaption}

\usepackage{bm}
\usepackage{url}
\usepackage{amsmath,amssymb}

\usepackage[top=30truemm,bottom=30truemm,left=25truemm,right=25truemm]{geometry}

\usepackage{graphicx}% Include figure files
\usepackage{dcolumn}% Align table columns on decimal point
\usepackage{bm}% bold math
\usepackage{hyperref}
\usepackage{booktabs}

\begin{document}

% \preprint{APS/123-QED}

\title{Online Harassment of Celebrities and Influencers}

\author{Masanori Takano\thanks{takano\_masanori@cyberagent.co.jp}~\thanks{Multi-disciplinary Information Science Center, CyberAgent, Inc.} \and Fumiaki Taka\thanks{Faculty of Sociology, Toyo University} \and Chiki Ogiue\thanks{Chiki Lab} \and Natsuki Nagata\thanks{Graduate School of Education, Hyogo University of Teacher Education}}

\date{ }% It is always \today, today,

\maketitle

% \keywords{Online Harassment, Celebrity, Influencer, Online Questionnaire Survey}

\begin{abstract}
Famous people, such as celebrities and influencers, are harassed online on a daily basis. Online harassment mentally disturbs them and negatively affects society. However, limited studies have been conducted on the online harassment victimization of famous people, and its effects remain unclear. We surveyed Japanese famous people ($N=213$), who were influential people who appeared on television and other traditional media and on social media, regarding online harassment victimization, emotional injury, and action against offenders and revealed that various forms of online harassment are prevalent. Some victims used the anti-harassment functions provided by weblogs and social media systems (e.g., blocking/muting/reporting offender accounts and closing comment forms), talked about their victimization to close people, and contacted relevant authorities to take legal action (talent agencies, legal consultants, and police). By contrast, some victims felt compelled to accept harassment and did not initiate action for offenses. We propose several approaches to support victims, inhibit online harassment, and educate people. Our findings help that platforms establish support systems against online harassment.
\end{abstract}

\section{Introduction}

Online harassment is a critical issue resulting from Internet use and considerably harms victims' mental health and dignity.
Online harassment includes cyberbullying, hate speech, flaming, doxing, impersonating, and public shaming~\cite{Blackwell2017}.
Famous people, such as celebrities and influencers, are harassed online on a daily basis~\cite{Bulck2014,Ouvrein2018,Hand2021}; however, the extent of injury and actions remains unclear because of the difficulties in conducting investigative surveys on famous people.
Most studies investigating the online harassment of celebrities are case studies using social media logs, such as ~\cite{Kim2013,Matamoros-Fernandez2017,Lawson2017,Lawson2018,Marwick2017,Park2021}.
Surveys of victims can provide an understanding of the online harassment of famous people.
For example, surveys of female journalists~\cite{Chen2018}, Japanese journalists~\cite{Yamaguchi2023a}, and politicians~\cite{Every-Palmer2015} have revealed victimization, mental and physical injuries, effects on their day-to-day activities (e.g., involuntarily inhibiting victims' activities), and actions against online harassment.
Our study reveals a quantitative picture of the online harassment victimization of famous people through a survey ($N=213$).
Accordingly, we discuss the mitigation of online harassment victimization.

The online harassment of famous people can easily become particularly extreme~\cite{Ouvrein2018,Scott2019,Lawson2017,Ouvrein2018,Lee2020} thereby harming victims~\cite{Bliuc2018} and can lead to suicide~\cite{Hinduja2010,Brailovskaia2018}.

Online harassment against famous people also causes indirect negative effects.
First, sensationalized coverage of the suicides of famous people increases suicide mimicry (Werther effect)~\cite{Phillips1974,Kim2013} because such coverage evokes increased negative emotions and feelings of social isolation than the coverage of suicides of other victims~\cite{Rosen2018}.
Second, observing online harassment encourages the aggressive behavior of observers~\cite{Bliuc2018,Ouvrein2018,Scott2019,Yokotani2021} and tempts them to justify their offensive behavior~\cite{Crandall2003}, e.g., they believe that celebrities are socially strong~\cite{Lawson2017,Ouvrein2018,Lee2020}.
Consequently, online harassment would escalate in intensity and frequency.
Third, spreading negative information that damages victims' reputations, such as negative gossip and disinformation, severely affects their businesses~\cite{Sarna2017}.
Fourth, the popularization of online harassment inhibits free expression and discussions among victims and other people~\cite{Chen2018,Yamaguchi2023a}.

Although famous people can be vulnerable to online harassment, they do not always act against it.
Victims may feel stigmatized by harassment and fear rejection by others when they speak out ~\cite{Mesch2006,Andalibi2016,Foster2018}.
A famous victim targeted by several people tends to attract victim-blaming~\cite{Hand2021}.
Furthermore, the victimization of famous people tends to attract sensational media coverage~\cite{Kang2022}.
Therefore, victims are often under severe suppression~\cite{Hand2021,Kang2022}.

Additionally, the anti-harassment actions taken by victims are often associated with several problems.
Offenders typically circumvent blocking/muting accounts by creating new ones.
The Internet is a vital tool for communication between famous people, such as celebrities and influencers, and their fans~\cite{Bulck2014}.
Therefore, closing comment forms decreases communication opportunities with fans and impairs business.
Actions against online harassment on other platforms (indirect harassment, e.g., harmful gossiping about victims) are limited~\cite{Sarna2017} and require complex legal procedures.
Therefore, many victims are forced to accept online harassment.

Social media platforms remove toxic comments and ban the accounts of people posting such comments to protect their users, including celebrities and influencers.
However, the efficacy of these provisions is limited.
Social media platforms cannot moderate all toxic comments~\cite{SaleemHMohammad2016ASpaces,Cho2017,Sap2019,Badjatiya2019,Milosevic2022}.
Linguistic filtering approaches can accidentally result in the biased treatment of discriminated minorities~\cite{Sap2019,Badjatiya2019}.
Removing toxic comments according to the terms of service only removes explicit expressions; therefore, ambiguous and/or cloaked blatant expressions tend to remain on platforms~\cite{SaleemHMohammad2016ASpaces,Cho2017,Milosevic2022}.
A study revealed that banning the accounts of offenders would not alter their thinking~\cite{Johnson2019}.

Online harassment victimization has some common characteristics across countries and cultures~\cite{Chen2018}. 
Famous Japanese people are also harassed online; e.g., 
Japanese journalists were more than four times more likely to experience online harassment than the general population~\cite{Yamaguchi2023a}; and 
Japanese professional wrestler Hana Kimura committed suicide owing to online harassment~\cite{Dooley2020}. 
A survey of Japanese Internet users conducted shortly after her suicide showed that the respondents tended to think that online harassment against famous people is the price of fame and/or unavoidable accidents~\cite{BIGLOBE2020}, as also found by previous studies in other countries~\cite{Lawson2017,Ouvrein2018,Lee2020}. 
A survey of Japanese journalists also indicate that the lack of organizational cover makes them easy targets for harassment, and that the number of people who take concrete action (e.g., blocking and talking family and police) when victimized is limited~\cite{Yamaguchi2023a}.
The lack of such actions has been noted in surveys of ordinary Japanese Internet users~\cite{Yamaguchi2023a}.
Additionally, the Japanese anonymous bulletin board 2channel, which is a toxic site on the Japanese Internet, has cultural roots in 4chan, which is a toxic site on the English Internet~\cite{Dooley2022}. 
Therefore, studying harmful online phenomena in Japan may have value beyond Japanese culture.

We study the following research questions:
\begin{itemize}
    \item {\bf RQ1}: What type of people are harassed online, and by whom?
    \item {\bf RQ2}: How do victims react to online harassment?
\end{itemize}
Based on the answers to these questions, we reveal the issues of the current online harassment victim support system and discuss ways to improve them.

The contributions of the study are as follows: 
\begin{itemize}
    \item A quantitative picture of the online harassment victimization of famous people is revealed.
    \item The characteristics of vulnerable famous people in online harassment are discussed.
    \item Issues with existing support systems against online harassment for famous people are identified.
    \item Based on these, we discuss approaches against online harassment.
\end{itemize}

\section{Dataset}

\subsection{Participants}

We conducted an online survey of famous people (September 27 to November 18, 2021). 
The participants in this survey were recruited by a Japanese Internet company, CyberAgent, Inc.\footnote{\url{https://www.cyberagent.co.jp/en/}}
They, who were business partners of CyberAgent, Inc., were influential people who appeared on television and other traditional media and on social media.
The number of potential participants was approximately 20,000.
The details of the population cannot be disclosed as they are a trade secret of CyberAgent Inc.
Participation in this survey is unpaid.
We analyzed the results of questionnaires that they fully answered ($N=213$). 
The participation rate of this survey was low; this may be because the survey was conducted without monetary compensation for respondents, and celebrities and influencers generally tend to have full schedules.
Note that these participants appeared to have participant biases because victims may tend to participate in this survey.

\subsection{Measures}

Our survey addressed victimization by online harassment.
We created questionnaire items by referring to a previous survey of harassment in Japan\footnote{\url{https://www.sra-chiki-lab.com/app/download/11636201821/\%E8\%A1\%A8\%E7\%8F\%BE\%E3\%81\%AE\%E7\%8F\%BE\%E5\%A0\%B4\%E3\%83\%8F\%E3\%83\%A9\%E3\%82\%B9\%E3\%83\%A1\%E3\%83\%B3\%E3\%83\%88\%E7\%99\%BD\%E6\%9B\%B8.pdf?t=1619570467}}.
Additionally, the participants' responses concerned emotional injury, actions against online harassment, victimization through offline harassment, demographic information, the number of online followers, and media appearance.
We investigated associations between them and victimization through online harassment.
% Detailed results of the exploratory factor analysis are available in the online supplement\footnote{\url{https://figshare.com/s/c5a0057739d5a8bca3f1}}.

\subsubsection{Direct and Indirect Online Harassment}

Offenders can directly attack victims online~\cite{Bulck2014,Sarna2017,Hand2021}. Online attacks include posting toxic comments to victims' weblogs, sending harmful replies to victims' social media accounts, and sending sexual messages to victims.
We investigated victimization through direct online harassment by different offender types using various channels.
Offender types were 1) individuals who repeatedly harassed victims (individual) and 2) the unspecified public (collective). 
The channels were public and private messages, e.g., public comments include comments on weblogs and replies on social media; private messages include direct messages on social media and e-mails.
We asked about victimization through direct online harassment according to the following activities:
my appearance has been abused (looks, body shape, etc.); 
my ability has been abused (knowledge, etc.); 
my personality has been abused; 
I received obscene messages (texts, images, etc.); I was threatened; 
and I have received requests for a sexual relationship.
We call them ``appearance abuse,'' ``ability abuse,'' ``personality abuse,'' ``obscene messages,'' ``intimidation,'' and ``requesting a sexual relationship,'' respectively.
The participants responded with ``yes,'' ``no,'' and ``I do not remember/want to respond.''
We evaluated the number of ``yes'' answers in our analyses.
In total, there were 24 questions (two offender types, two channels, and six victimization types).

Offenders can also harass victims indirectly~\cite{Sarna2017} through activities such as harmful gossiping about victims in other places, disclosing victims' personal information, sharing disinformation, and uploading photoshopped sexual images.
We asked about victimization through indirect online harassment according to the following criteria:
my photos and videos were uploaded without authorization; 
my sexually suggestive photos and videos were uploaded, for example, zooming necklines and underclothes;
my ungainly photographs and videos were uploaded, for example, rolling their eyes up into their heads;
my photoshopped photos and videos were uploaded;
my personal information was disclosed, for example, my residential address;
I assisted in situations that were difficult for them to respond to, for example, suicidal feelings;
my disinformation was spread on social media;
and my disinformation was published on curation and news sites.
Participants responded with ``yes,'' ``no,'' or ``I do not remember/want to respond.''
We evaluated the number of ``yes'' answers in our analyses.
There were eight questions.

\begin{table*}[t!]																			
\centering						
\footnotesize
% \hspace{-6.5cm}
  \begin{tabular}{ll|rrrrr}
    \toprule
Category	&	Item	&		F1		&		F3		&		F2		&		F4		&		F5		\\ \midrule
Direct (public-individual)	&	My personality has been abused.	&	{\bf	0.81	}	&		0.28		&		-0.10		&		-0.08		&		-0.20		\\
Direct (public-collective)	&	My appearance has been abused.	&	{\bf	0.79	}	&		-0.10		&		0.08		&		0.24		&		-0.17		\\
Direct (public-individual)	&	My appearance has been abused.	&	{\bf	0.79	}	&		-0.18		&		0.00		&		0.33		&		-0.12		\\
Direct (public-collective)	&	My personality has been abused.	&	{\bf	0.75	}	&		0.31		&		0.07		&		-0.15		&		-0.24		\\
Direct (public-individual)	&	My ability has been abused.	&	{\bf	0.57	}	&		0.24		&		-0.09		&		-0.03		&		0.08		\\
Direct (public-collective)	&	My ability has been abused.	&	{\bf	0.54	}	&		0.43		&		-0.02		&		-0.14		&		0.04		\\
Indirect	&	My disinformation was spread on social media.	&	{\bf	0.49	}	&		0.17		&		0.01		&		-0.04		&		0.08		\\
Indirect	&	My photoshopped photos and videos were uploaded.	&	{\bf	0.48	}	&		-0.20		&		0.21		&		0.03		&		0.17		\\
Direct (public-collective)	&	I received obscene messages.	&	{\bf	0.43	}	&		0.00		&		-0.07		&		0.38		&		-0.15		\\
Indirect	&	My ungainly photographs and videos were uploaded.	&	{\bf	0.40	}	&		-0.01		&		0.16		&		0.28		&		-0.02		\\
Indirect	&	My disinformation was published on curation and news sites.	&	{\bf	0.39	}	&		0.04		&		0.13		&		0.03		&		0.13		\\
Indirect	&	My personal information was disclosed.	&	{\bf	0.33	}	&		-0.08		&		0.27		&		0.00		&		-0.02		\\
Indirect	&	My sexually suggestive photos and videos were uploaded.	&	{\bf	0.16	}	&		-0.07		&		-0.13		&		0.06		&		0.09		\\
Direct (private-collective)	&	My personality has been abused.	&		0.06		&	{\bf	0.82	}	&		0.04		&		0.17		&		-0.16		\\
Direct (private-individual)	&	My personality has been abused.	&		0.02		&	{\bf	0.81	}	&		-0.04		&		0.29		&		-0.11		\\
Direct (private-collective)	&	My ability has been abused.	&		-0.05		&	{\bf	0.77	}	&		0.00		&		0.16		&		0.14		\\
Direct (private-individual)	&	My ability has been abused.	&		-0.08		&	{\bf	0.74	}	&		-0.06		&		0.27		&		0.19		\\
Indirect	&	I assisted in situations that were difficult for them to respond to.	&		0.16		&	{\bf	0.25	}	&		-0.01		&		-0.25		&		0.12		\\
Direct (private-collective)	&	I was threatened.	&		-0.07		&		-0.01		&	{\bf	0.92	}	&		0.02		&		0.13		\\
Direct (public-collective)	&	I was threatened.	&		0.02		&		0.05		&	{\bf	0.87	}	&		0.03		&		-0.06		\\
Direct (private-individual)	&	I was threatened.	&		-0.05		&		-0.07		&	{\bf	0.86	}	&		0.07		&		0.14		\\
Direct (public-individual)	&	I was threatened.	&		0.12		&		-0.01		&	{\bf	0.74	}	&		-0.03		&		-0.04		\\
Indirect	&	My photos and videos were uploaded without authorization.	&		0.12		&		0.10		&	{\bf	0.16	}	&		0.16		&		0.12		\\
Direct (private-individual)	&	My appearance has been abused.	&		0.01		&		0.18		&		0.06		&	{\bf	0.85	}	&		-0.01		\\
Direct (private-collective)	&	My appearance has been abused.	&		-0.02		&		0.17		&		0.05		&	{\bf	0.73	}	&		0.06		\\
Direct (private-individual)	&	I received obscene messages.	&		-0.03		&		0.09		&		0.02		&	{\bf	0.57	}	&		0.09		\\
Direct (private-collective)	&	I received obscene messages.	&		0.07		&		0.05		&		0.02		&	{\bf	0.44	}	&		-0.02		\\
Direct (public-individual)	&	I received obscene messages.	&		0.40		&		0.01		&		-0.14		&	{\bf	0.41	}	&		-0.01		\\
Direct (private-individual)	&	I have received requests for a sexual relationship.	&		-0.20		&		0.04		&		-0.01		&		0.13		&	{\bf	0.84	}	\\
Direct (public-collective)	&	I have received requests for a sexual relationship.	&		-0.16		&		0.08		&		0.02		&		0.07		&	{\bf	0.80	}	\\
Direct (public-individual)	&	I have received requests for a sexual relationshipv	&		0.08		&		-0.06		&		0.07		&		-0.13		&	{\bf	0.78	}	\\
Direct (public-collective)	&	I have received requests for a sexual relationship.	&		0.03		&		0.02		&		0.05		&		-0.02		&	{\bf	0.72	}	\\ \midrule
	&	Factor correlations	&		F1		&		0.45		&		0.41		&		0.48		&		0.39		\\
	&		&		F3		&				&		0.39		&		0.19		&		0.35		\\
	&		&		F2		&				&				&		0.26		&		0.38		\\
	&		&		F4		&				&				&				&		0.24		\\ \bottomrule
    \end{tabular}
\caption{Result of the EFA of direct and indirect online harassment
}
  \label{tab_efa_harassment}
\end{table*}	

Exploratory factor analysis (EFA) revealed five factors for these 32 items by using maximum likelihood estimation (MLE) and Promax rotation. We selected the number of factors using the Bayesian information criterion (BIC).
The comparative fit index (CFI)~\cite{Bentler1990} was $0.800$ and the root mean square error of approximation (RMSEA)~\cite{Steiger1990} was $0.11~[0.103, 0.117]$, where the square brackets indicate a 90\% confidence interval.
These factors were interpreted as harassment in public spaces, harassment in private spaces, intimidation, dating requests, and sexual harassment (Table~\ref{tab_efa_harassment} in the supplementary materials).

\subsubsection{Emotional Injury}

Online harassment injures victims' mental health~\cite{Bliuc2018}.
To measure such injury, we assessed emotional injury due to online harassment using the following prompts:
I wanted to remove my account;
I did not want to use the Internet;
I felt less motivated to go out and interact socially;
I had a sinking feeling for several days;
I wanted to be truant or leave my job;
I became suicidally depressed;
I indulged in substance abuse;
and I engaged in self-harm and wrecked some things.
Participants responded based on a 4-point Likert-type scale, i.e., agree, slightly agree, slightly disagree, and disagree.
There were eight questions.

\begin{table*}[h!]																			
\centering						
% \renewcommand{\arraystretch}{0.932}
% \footnotesize
  \begin{tabular}{l|r}
    \toprule
Item	&	F1	\\ \midrule
I did not want to use the Internet.	&	0.83	\\
I wanted to remove my account.	&	0.83	\\
I felt less motivated to go out and interact socially.	&	0.81	\\
I had a sinking feeling for several days.	&	0.80	\\
I wanted to be truant or leave my job.	&	0.79	\\
I became suicidally depressed.	&	0.69	\\
I indulged in substance abuse.	&	0.52	\\
I engaged in self-harm and wrecked some things.	&	0.37	\\ \bottomrule
    \end{tabular}
\caption{Result of the EFA of emotional injury
}
  \label{tab_efa_damage}
\end{table*}

Using EFA, we acquired one factor, i.e., emotional injury (Table~\ref{tab_efa_damage} in the supplementary materials), for these eight items by applying MLE and Promax rotation, from which we selected the number of factors using the BIC (CFI: $0.923$, RMSEA: $0.128~[0.102, 0.156]$).

\subsubsection{Actions against Online Harassment}

Victims can initiate several actions against online harassment, including talking to authorities and asking platforms for restrictions on and punishment of offender accounts~\cite{Every-Palmer2015}.
We investigated the actions taken against online harassment with the following prompts:
I avoided and consigned weblog/social media posts to my staff;
 I restricted comments and replies on my weblog/social media;
 I blocked/muted harasser accounts on platforms;
 I reported harassers to platforms;
 I stopped updating my weblog/social media;
 I contacted an inquiry counter of the platforms;
 I talked family/friends/business friends;
 I reported my victimization to my talent agency;
 I went to mental health counselors;
 I discussed with my legal consultants;
 and I spoke to the police.
Participants responded with ``yes,'' ``no,'' or ``I do not remember/want to respond.''
We evaluated the number of ``yes'' answers in our analyses.
There were 11 questions.

\begin{table*}[h!]																			
\centering						
% \renewcommand{\arraystretch}{0.932}
%  \tabcolsep=2pt
% \footnotesize
  \begin{tabular}{l|rr}
    \toprule
Item	& F2	&	F1	\\ \midrule
I discussed with my legal consultants.&	0.70	&	-0.08	\\
I spoke to the police.	&0.67	&	-0.12	\\
I went to mental health counselors.	&0.49	&	-0.05	\\
I contacted an inquiry counter of the platforms.&	0.42	&	0.09	\\
I reported my victimization to my talent agency. &	0.41	&	-0.01	\\
I talked family/friends/business friends.&	0.27	&	0.10	\\
I avoided and consigned weblog/social media posts to my staff.&	0.22	&	0.02	\\
I blocked/muted harasser accounts on platforms.&	-0.08	&	0.90	\\
I restricted comments and replies on my weblog/social media.&	-0.15	&	0.65	\\
I reported harassers to platforms.	&0.07	&	0.61	\\
I stopped updating my weblog/social media.	&0.05	&	0.14	\\ \midrule
Factor correlations&	F2	&	0.47	\\
    \end{tabular}
\caption{Result of the EFA of actions against online harassment
}
  \label{tab_efa_action}
\end{table*}

Through EFA, we acquired two factors for these 11 items by applying MLE and Promax rotation, from which the number of factors was selected using the BIC (CFI: $0.956$, RMSEA: $0.043~[0.000, 0.070]$).
These factors were understood to discuss victimization and the use of the functions provided by platforms (Table~\ref{tab_efa_action} in the supplementary materials).

\subsubsection{Offline Harassment}

Victims of online harassment are more likely to experience victimization in the real world~\cite{Every-Palmer2015,Meloy2016}.
We assessed offline victimization of stalking based on the following:
 I received persistent requests for dating and companionship;
 I was stalked, ambushed, and intruded on;
 I was followed by a stranger;
 I was informed that I was being monitored;
 A GPS device was attached or tracking applications were installed without any prior consent;
 I uninterruptedly received calls, faxes, and e-mails;
 I received unwanted letters and/or presents;
 I was sent body fluid and/or excrement;
 and I was bugged, and/or I received a present with a bugging device.
Participants responded with ``yes,'' ``no,'' or ``I don't remember/want to respond.''
In our analyses, we evaluated the number of ``yes'' answers.

There were nine questions.
Note that no respondents with ``yes'' on the fifth, eighth, and ninth items.
Therefore, we analyzed the other six items.

\begin{table*}[h!]																			
\centering						
% \renewcommand{\arraystretch}{0.932}
% \footnotesize
  \begin{tabular}{l|r}
    \toprule
Item	&		F1		\\ \midrule
I was stalked, ambushed, and intruded on.	&	0.87	\\
I was followed by a stranger.	&	0.79	\\
I received persistent requests for dating and companionship.	&	0.68	\\
I received unwanted letters and/or presents.	&	0.66	\\
I was informed that I was being monitored.	&	0.60	\\
I uninterruptedly received calls, faxes, and e-mails.	&	0.50	\\ \bottomrule
    \end{tabular}
\caption{Result of the EFA of offline harassment
}
  \label{tab_efa_stalking}
\end{table*}

Using EFA, we acquired one factor, offline stalking (Table~\ref{tab_efa_stalking} in the supplementary materials), for these six items to which we applied MLE and Promax rotation and from which we selected the number of factors using the BIC (CFI: $0.962$, RMSEA: $0.098~[0.057, 0.141]$).

\subsubsection{Demographic Information}
The participants provided their demographic information, including gender (female, male, and ``I do not respond'') and age in ten-year intervals  (under 20\footnote{We instructed participants under 15 years to respond with their parents or guardians.}; 20s, 30s, 40s, 50s, and over 60s; and ``I do not respond'').
We used this coarse age-level grading to avoid identifying famous people about whom much information is publicly available.
In the following regression analyses, ``male'' is a reference category for gender.
The numbers of women, men, and no response were 172, 39, and 2, respectively.
The numbers of participants in the age levels from the under 20s, 20s, 30s, 40s, 50s, and over 60s groups were 0, 2, 57, 91, 41, and 6, respectively.
``I do not respond.'' was selected 16 times.

\subsubsection{Number of Online Followers}
The participants provided the total number of their online followers (weblogs, social media, video streaming services, etc.) in four levels (5,000 or less, 5,001 to 10,000, 10,001 to 100,000, and over 100,000).
The numbers of these levels were 62, 36, 92, and 23, respectively.

\subsubsection{Media Appearance}
The participants detailed how much exposure they experienced on conventional and online media e.g., TV, radio, newspapers, magazines, online video media, and online text media. This was based on 4-point Likert-type scales, i.e., high, moderate, low, and none.

\subsection{Statistical Model}

\begin{figure}[t!]
\centering
  \includegraphics[width=1\columnwidth]{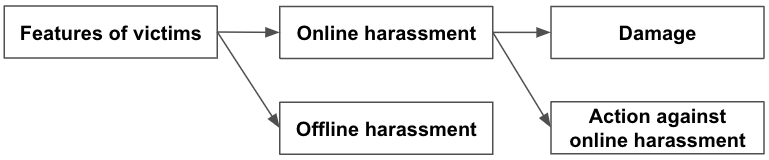}
   \caption{Overview of the SEM}
  ~\label{fig_sem_model}
\end{figure}

To detail the associations between victims' features (demographic information, number of online followers, media exposure), direct and indirect victimization, offline stalking, emotional injury, and actions against harassment, we analyzed a structural equation model (SEM; Fig~\ref{fig_sem_model}).
We considered demographic information, the number of followers (fame and influence on social media), and mass/online media exposure (fame depending on media types) as victim features. The relationships between these victim features and online harassment victimization can explain who is vulnerable and how. The victimization should have damaged the victims emotionally and they did or did not act against the harassment. Therefore, we analyzed the associations between each victim's features, victimization, emotional injury, and actions against harassment.
We used the EFA results for the variables of direct and indirect victimization, offline stalking, emotional injury, and actions against harassment.
We excluded the participants who did not respond to gender and/or age from this analysis.
The number of participants in this analysis was $197$.
We used age levels as an ordinal measure with normalization.

This model was used to analyze the types of people vulnerable to particular types of online harassment, the types of online harassment harmful to victims' mental health, and the actions initiated by victims.
Offline and online harassment were examined and compared.

\subsection{Ethics Statement}

% 倫理審査こちらです
% 研究代表者	高野雅典
% 所属	株式会社サイバーエージェント
% 研究題目	日本の著名人・インフルエンサーに対するオンラインハラスメントの計量調査
% 申請番号	P230067
% 受理日	2024/1/12
% 審査結果	承認
% 付帯事項 なし

% 承認機関は東洋大学社会学研究科倫理審査委員会
Our study was approved by the ethics committee of Graduate School of Sociology, Toyo University (P230067). 
Furthermore, all procedures were conducted in accordance with the guidelines for studies involving human participants, the ethical standards of the institutional research committee, and the 1964 Helsinki Declaration and its later amendments. 
Informed consent was obtained from the participants for participation in our survey, and the purpose of this study was fully explained.
The participants allowed the authors to analyze their data and anti-harassment actions for academic purposes on condition that the quantitative data outputs were aggregated, meaning no identifying information was presented.

CyberAgent, Inc. facilitated this questionnaire survey based on its privacy policy\footnote{\url{https://www.cyberagent.co.jp/way/security/privacy/}} and the informed consent for the survey.
The survey was anonymized.

\section{Results}

\subsection{Online Harassment Victimization}

\begin{figure}[t]
\centering
  \includegraphics[width=1\columnwidth]{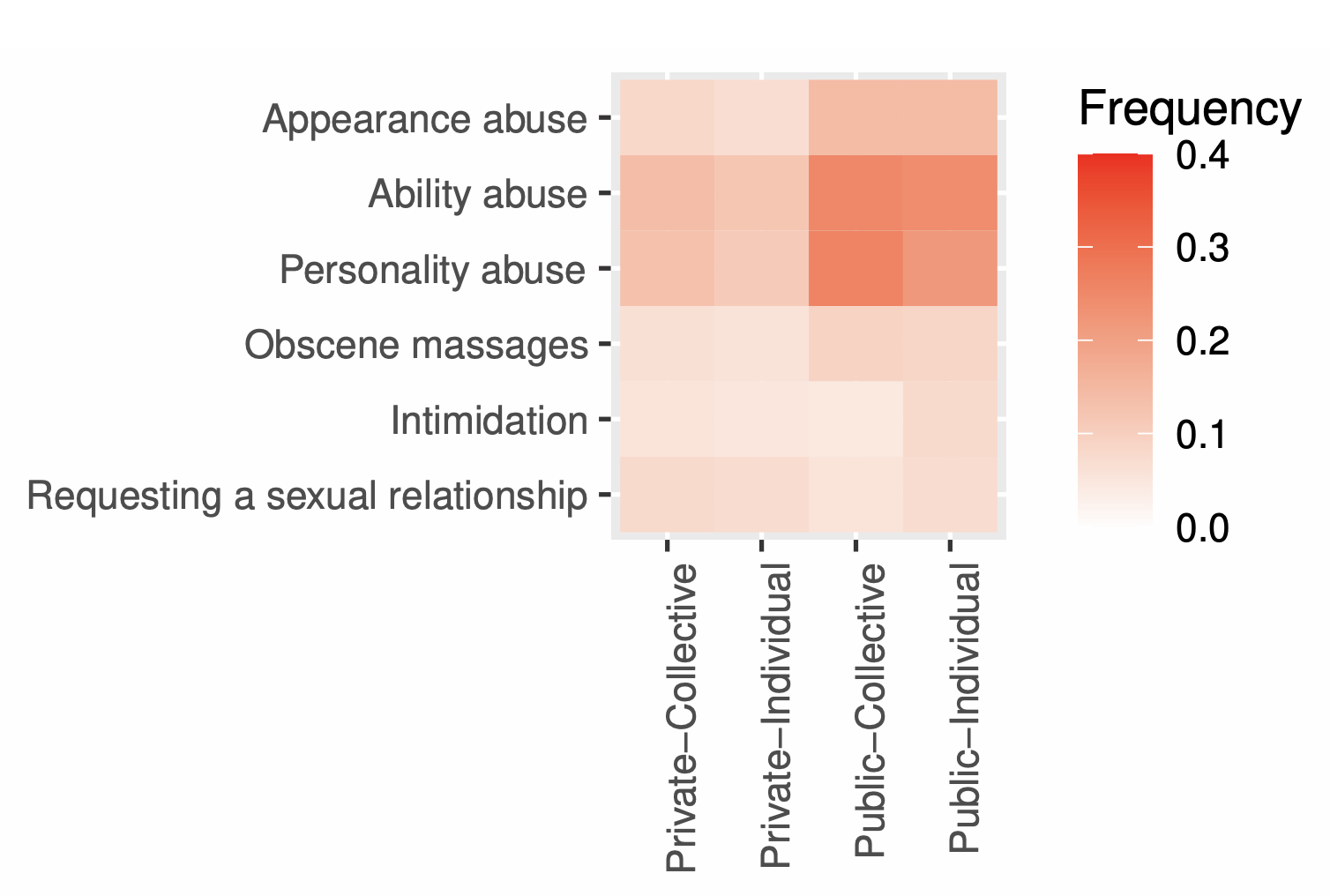}
   \caption{Direct harassment frequencies of offender types and methods}~\label{fig_dhrs}
\end{figure}
\begin{table*}[t!]																			
\centering						
% \renewcommand{\arraystretch}{0.932}
% \footnotesize
% \hspace{-6.5cm}
  \begin{tabular}{l|rrrr}
    \toprule
Harassment type &	Private-Collective	&	Private-Individual	&	Public-Collective	&	Public-Individual	\\ \midrule
Appearance abuse	&	0.08	&	0.07	&	0.15	&	0.15	\\
Ability abuse	&	0.14	&	0.12	&	0.25	&	0.24	\\
Personality abuse	&	0.13	&	0.11	&	0.26	&	0.22	\\
Obscene massages	&	0.07	&	0.06	&	0.09	&	0.09	\\
Intimidation	&	0.06	&	0.05	&	0.05	&	0.08	\\
Requesting a sexual relationship	&	0.08	&	0.08	&	0.06	&	0.08	\\ \bottomrule
    \end{tabular}
\caption{Direct harassment frequencies of offender types and methods}~\label{tbl_dhrs}
\end{table*}

Fig.~\ref{fig_dhrs} displays the frequencies of direct harassment according to offender types and channels (see Table~\ref{tbl_dhrs} for details).
Offenders resorted to direct harassment in public forums (the victimization ratio was 43\%), such as replying to social media and weblog comments.
Most of these messages were abusive about physical or psychological characteristics.
Victims tended to be targeted by both types of offenders, i.e., particular repeat offenders (individual) and unspecified masses (collective).
Toxic messages were conveyed to victims also through private messages (the victimization ratio was 27\%).
Forty-nine percent of the participants reported being subjected to at least one type of direct harassment.

\begin{figure}[t]
\centering
  \includegraphics[width=1\columnwidth]{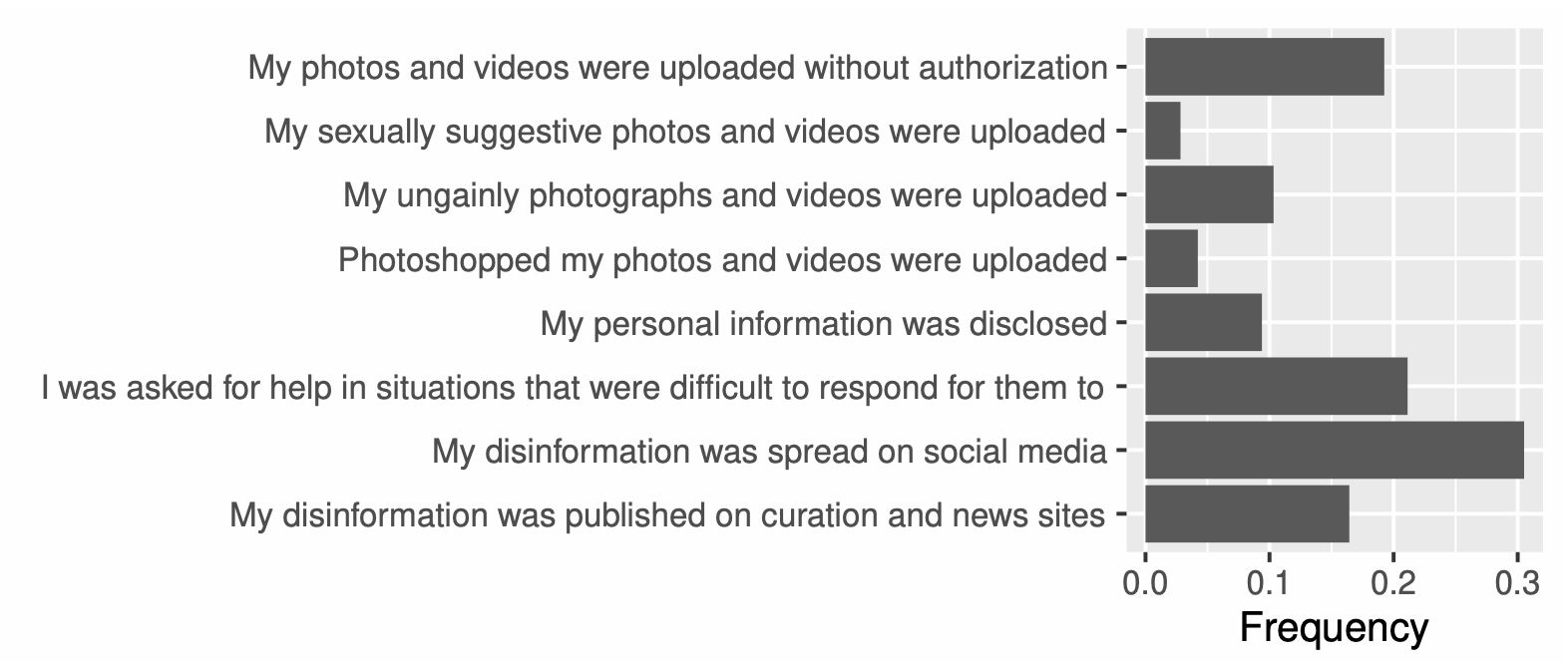}
   \caption{Indirect harassment frequencies}~\label{fig_indhrs}
\end{figure}

\begin{table*}[t!]																			
\centering						
% \footnotesize
% \hspace{-6.5cm}
  \begin{tabular}{l|r}
    \toprule
Harassment type 	&	Frequency	\\ \midrule
My photos and videos were uploaded without authorization	&	0.19	\\
My sexually suggestive photos and videos were uploaded	&	0.03	\\
My ungainly photographs and videos were uploaded	&	0.10	\\
Photoshopped my photos and videos were uploaded	&	0.04	\\
My personal information was disclosed	&	0.09	\\
I was asked for help in situations that were difficult to respond for them to	&	0.21	\\
My disinformation was spread on social media	&	0.31	\\
My disinformation was published on curation and news sites	&	0.16	\\
\bottomrule
    \end{tabular}
\caption{Indirect harassment frequencies}~\label{tbl_indhrs}
\end{table*}

Fig.~\ref{fig_indhrs} displays the frequencies of indirect harassment (see Table~\ref{tbl_indhrs} for details).
The participants were exposed to indirect harassment, such as negative gossip, and disinformation (spread on social media and published on curation and news sites).
Offenders shamed victims with unflattering, sexual, and photoshopped images and disclosed victims' personal information.
Additionally, the victims were sometimes required to assist by answering questions that were difficult to respond to, e.g., suicidal feelings.
Forty-nine percent of the participants reported being subjected to at least one type of indirect harassment.

\begin{figure}[t]
\centering
  \includegraphics[width=1\columnwidth]{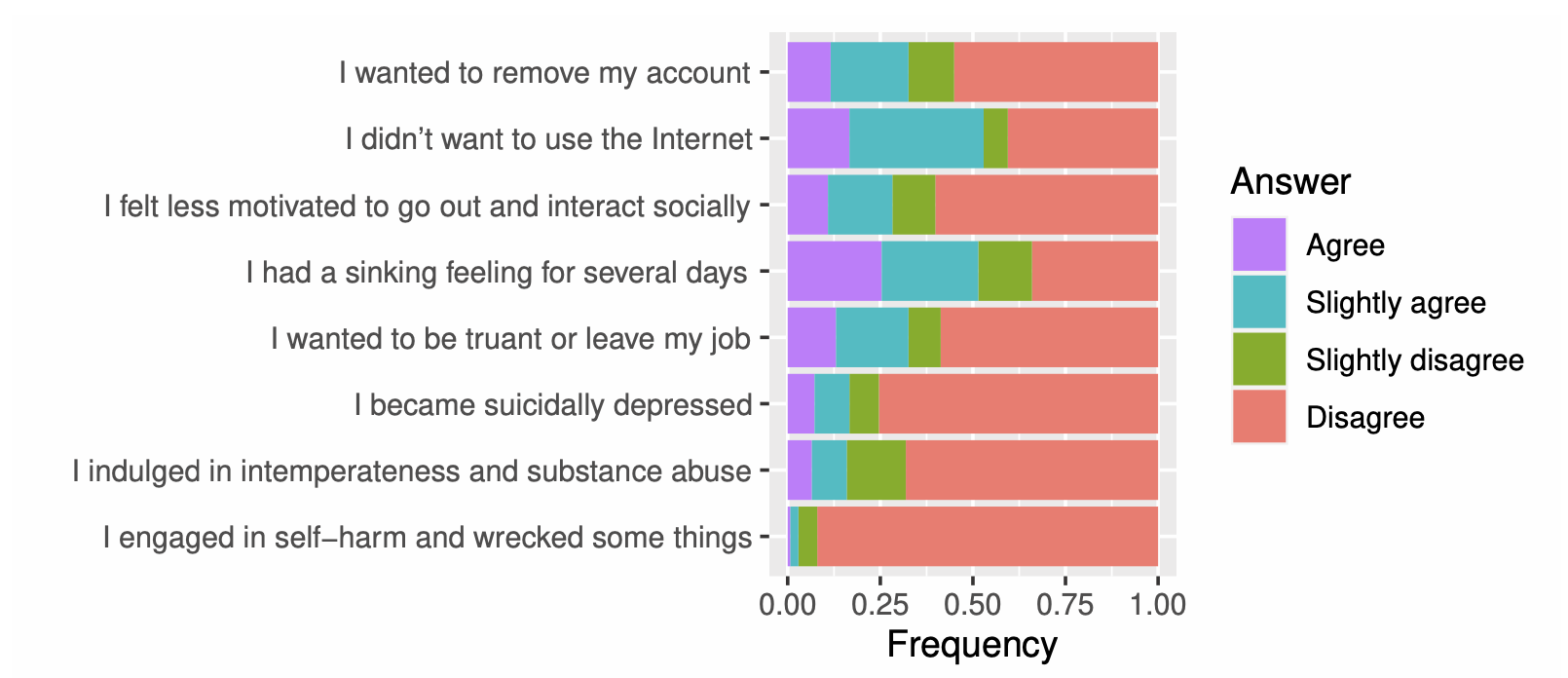}
   \caption{Emotional damages}
  ~\label{fig_damage}
\end{figure}

\begin{table*}[t!]
\centering
% \footnotesize
% \hspace{-6.5cm}
\begin{tabular}{l|rrrr}
\toprule
Emotional damage & Disagree & Slightly disagree & Slightly agree & Agree \\ \midrule
I wanted to remove my account & 0.55 & 0.12 & 0.21 & 0.12 \\ 
I didn't want to use the Internet & 0.41 & 0.07 & 0.36 & 0.17\\
I felt less motivated to go out and interact socially & 0.6 & 0.12 & 0.17 & 0.11 \\
I had a sinking feeling for several days & 0.34 & 0.14 & 0.26 & 0.25 \\
I wanted to be truant or leave my job & 0.59 & 0.09 & 0.20 & 0.13 \\
I became suicidally depressed & 0.75 & 0.08 & 0.09 & 0.07 \\ 
I indulged in intemperateness and substance abuse & 0.68 & 0.16 & 0.09 & 0.07 \\
I engaged in self-harm and wrecked some things & 0.92 & 0.05 & 0.02 & 0.01 \\ \bottomrule
\end{tabular}
\caption{Emotional damages}~\label{tbl_damage}
\end{table*}

\begin{figure}[t]
\centering
  \includegraphics[width=1\columnwidth]{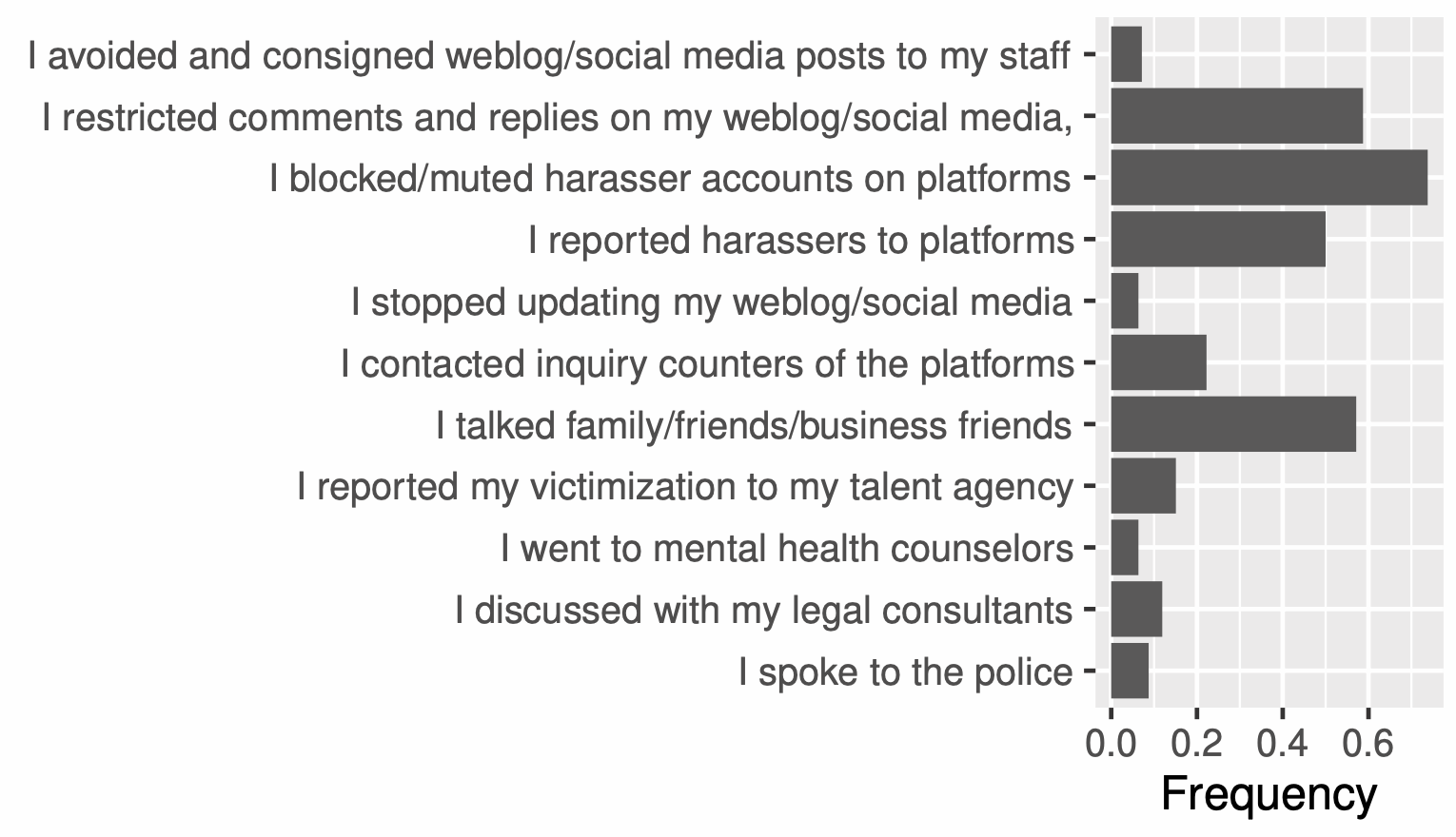}
   \caption{Action against online harassment}
  ~\label{fig_action}
\end{figure}

\begin{table*}[t!]
\centering
% \footnotesize
% \hspace{-6.5cm}
\begin{tabular}{l|r}
\toprule
Action & Frequency \\ \midrule
I avoided and consigned weblog/social media posts to my staff	&	0.07	\\
I restricted comments and replies on my weblog/social media,	&	0.59	\\
I blocked/muted harasser accounts on platforms	&	0.74	\\
I reported harassers to platforms	&	0.5	\\
I stopped updating my weblog/social media	&	0.06	\\
I contacted inquiry counters of the platforms	&	0.22	\\
I talked family/friends/business friends	&	0.57	\\
I reported my victimization to my talent agency	&	0.15	\\
I went to mental health counselors	&	0.06	\\
I discussed with my legal consultants	&	0.12	\\
I spoke to the police	&	0.09	\\ \bottomrule
\end{tabular}
   \caption{Action against online harassment}
  ~\label{tbl_action}
\end{table*}

\begin{figure*}[t]
\centering\includegraphics[width=1\columnwidth]{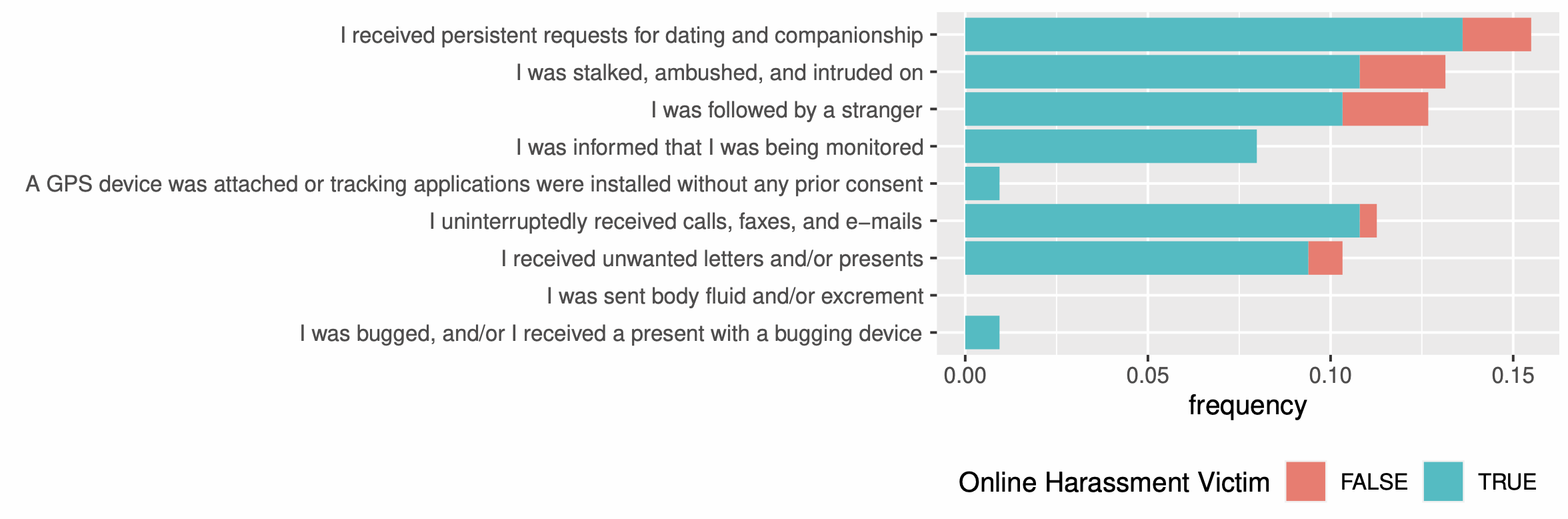}
   \caption{Association between offline stalkings and online harassment}
  ~\label{fig_offline}
\end{figure*}

\begin{table*}[t!]
\centering
\footnotesize
% \hspace{-7.5cm}
\begin{tabular}{l|rr}
\toprule
Offline harassment	&	Online harassment victim	&	Others	\\ \midrule
I received persistent requests for dating and companionship	&	0.14	&	0.02	\\
I was stalked, ambushed, and intruded on	&	0.11	&	0.02	\\
I was followed by a stranger	&	0.10	&	0.02	\\
I was informed that I was being monitored	&	0.08	&	0.00	\\
A GPS device was attached or tracking applications were installed without any prior consent	&	0.01	&	0.00	\\
I uninterruptedly received calls, faxes, and e-mails	&	0.11	&	0.00	\\
I received unwanted letters and/or presents	&	0.09	&	0.01	\\
I was sent body fluid and/or excrement	&	0.00	&	0.00	\\
I was bugged, and/or I received a present with a bugging device	&	0.01	&	0.00	\\ \bottomrule
\end{tabular}
   \caption{Association between offline stalkings and online harassment}
  ~\label{tbl_offline}
\end{table*}

Victims felt that online harassment negatively affected their mental health (Fig.~\ref{fig_damage}; see Table~\ref{tbl_damage} for details).
The harassment resulted in them avoiding using the Internet (53\%), caused ``sinking feelings'' (51\%), led them to almost remove their accounts (33\%), and instilled the desire to skip or leave their work (33\%).
The values in parentheses are ``agree and slightly agree'' answer ratios of the participants.
Twenty-eight percent of victims complained of intense mental injury (suicidal feelings, self-harm reckless behavior, and intemperance/substance abuse.

Victims often used anti-harassment functions provided by platforms (Fig.~\ref{fig_action}; see Table~\ref{tbl_action} for details), such as blocking/muting offender accounts, closing comment forms, restricting accounts from posting comments, and reporting offender accounts to platforms to ban offenders.

Victims talked about their victimization to close people (family, friends, and business friends).
They also spoke to their associated organizations, i.e., talent agencies and platforms \footnote{Normally, several platforms seem to work with famous people.}.
A small number of victims consulted the police and legal counsel.

A few victims stopped their weblog/social media activities completely. 

Online harassment victims were also stalked offline (Fig.~\ref{fig_offline}; see Table~\ref{tbl_offline} for details).
Victims of online harassment were 3.1 times more likely to be stalked than those who were not.

\subsection{Vulnerable People, Emotional Injury, and Actions Against Online Harassment}

\begin{figure*}[t]
\centering
  \includegraphics[width=1.0\columnwidth]{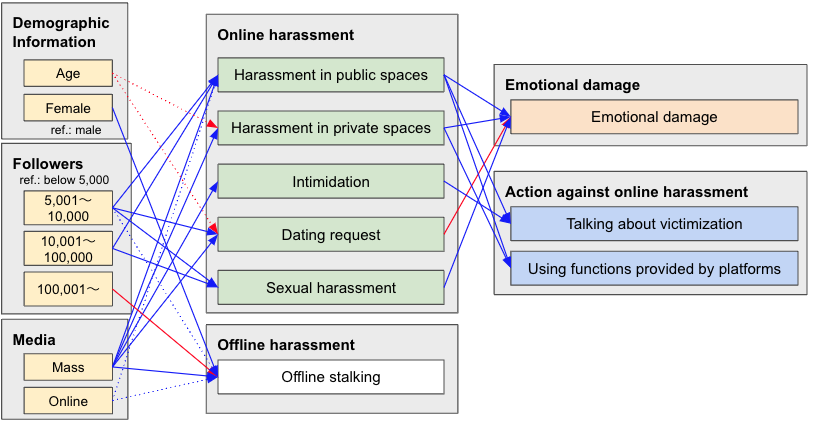}
   \caption{
   The results of SEM for the relationships between vulnerable people, damages, and action (CFI: $0.993$; RMSEA: $0.030~[0.000, 0.065]$).
 We only show paths which show statistically significant (solid arrows: $p\mathrm{-value} < 0.05$; dotted arrows: $0.05 \leq p\mathrm{-value} < 0.1$).
   The gray boxes detail the categories of the variables.
   ``ref.'' indicates a reference variable.
   The blue and red arrows express positive and negative associations, respectively.}
~\label{fig_sem}
\end{figure*}

\begin{table}[t!]																			
\centering						
\renewcommand{\arraystretch}{0.922}
\small
\tabcolsep = 2pt
  \begin{tabular}{l|l|r|l}
    \toprule
Explained variable	&	Explanatory variable	&	Coef.	&	S.E.		\\ \midrule
Harassment 	&	Female	&	0.17	&	0.13		\\
~in public spaces	&	Age	&	-0.01	&	0.06		\\
	&	Followers: 5,001-10,000	&	0.50	&	0.20	*	\\
	&	Followers: 10,001-100,000	&	0.32	&	0.13	*	\\
	&	Followers: 100,001-	&	0.26	&	0.22		\\
	&	Mass media	&	0.28	&	0.10	**	\\
	&	Online media	&	0.16	&	0.09	.	\\ \midrule
Harassment 	&	Female	&	0.08	&	0.15		\\
~in private spaces	&	Age	&	-0.11	&	0.06	.	\\
	&	Followers: 5,001-10,000	&	0.08	&	0.20		\\
	&	Followers: 10,001-100,000	&	0.18	&	0.17		\\
	&	Followers: 100,001-	&	-0.21	&	0.23		\\
	&	Mass media	&	0.34	&	0.10	***	\\
	&	Online media	&	-0.01	&	0.09		\\ \midrule
Intimidation	&	Female	&	-0.27	&	0.21		\\
	&	Age	&	0.04	&	0.06		\\
	&	Followers: 5,001-10,000	&	0.39	&	0.24		\\
	&	Followers: 10,001-100,000	&	0.14	&	0.13		\\
	&	Followers: 100,001-	&	0.23	&	0.33		\\
	&	Mass media	&	0.32	&	0.16	*	\\
	&	Online media	&	-0.10	&	0.12		\\ \midrule
Dating request	&	Female	&	-0.16	&	0.20		\\
	&	Age	&	-0.11	&	0.06	.	\\
	&	Followers: 5,001-10,000	&	0.66	&	0.28	*	\\
	&	Followers: 10,001-100,000	&	0.02	&	0.11		\\
	&	Followers: 100,001-	&	0.06	&	0.24		\\
	&	Mass media	&	0.24	&	0.10	**	\\
	&	Online media	&	-0.05	&	0.08		\\ \midrule
Sexual harassment	&	Female	&	0.13	&	0.16		\\
	&	Age	&	-0.02	&	0.06		\\
	&	Followers: 5,001-10,000	&	0.50	&	0.21	*	\\
	&	Followers: 10,001-100,000	&	0.26	&	0.12	*	\\
	&	Followers: 100,001-	&	0.38	&	0.32		\\
	&	Mass media	&	0.14	&	0.11		\\
	&	Online media	&	0.04	&	0.09		\\ \midrule
Offline stalking	&	Female	&	0.42	&	0.13	***	\\
	&	Age	&	0.02	&	0.06		\\
	&	Followers: 5,001-10,000	&	0.43	&	0.23	.	\\
	&	Followers: 10,001-100,000	&	-0.18	&	0.13		\\
	&	Followers: 100,001-	&	-0.40	&	0.19	*	\\
	&	Mass media	&	0.21	&	0.10	*	\\
	&	Online media	&	0.18	&	0.10	.	\\ \midrule
Emotional injury	&	Harassment in public spaces	&	0.27	&	0.10	**	\\
	&	Harassment in private spaces	&	0.20	&	0.08	*	\\
	&	Intimidation	&	0.06	&	0.07		\\
	&	Dating request	&	-0.16	&	0.07	*	\\
	&	Sexual harassment	&	0.17	&	0.08	*	\\ \midrule
Talking 	&	Harassment in public spaces	&	0.49	&	0.12	***	\\
~about victimization	&	Harassment in private spaces	&	-0.05	&	0.06		\\
	&	Intimidation	&	0.26	&	0.08	***	\\
	&	Dating request	&	0.08	&	0.08		\\
	&	Sexual harassment	&	-0.04	&	0.08		\\ \midrule
Using functions 	&	Harassment in public spaces	&	0.34	&	0.07	***	\\
~provided by platforms	&	Harassment in private spaces	&	0.22	&	0.05	***	\\
	&	Intimidation	&	-0.06	&	0.06		\\
	&	Dating request	&	0.05	&	0.06		\\
	&	Sexual harassment	&	0.00	&	0.05		\\ \bottomrule 
\end{tabular}
\caption{Details of the SEM result.
$^{***}$, $^{**}$, $^{*}$, and $^.$ indicate significant differences at $p\mathrm{-values} \leq 0.001, 0.01, 0.05, \mathrm{and}~0.1$.
}
  \label{tab_sem}
\end{table}		

Fig.~\ref{fig_sem} and Table~\ref{tab_sem} show the results of the SEM.

People with a modest number of followers are targeted for harassment in public spaces, sent dating requests, and sexually harassed.
This phenomenon could be attributed to offenders not finding opportunities to interact with people with a few followers and hesitating to attack people with many followers because such people wield tremendous influence on the Internet.
Offline harassment showed similar tendencies to online harassment.

People with many mass media appearances were targets of all types of online harassment, excluding sexual harassment.
By contrast, online media appearances did not reveal considerable effects, excluding harassment in public spaces.

Younger people may be targeted for harassment in private spaces and with dating requests.
There were insignificant gender differences in online harassment.
Although offline harassment followed online harassment trends, women were more vulnerable to offline stalking than men.

Victims of sexual harassment and harassment in public/private spaces were emotionally damaged.
Reported experiences of undesirable dating requests did not cause the victims severe injury.
This seems to be because persistent and vulgar dating requests could be considered sexual harassment.
Intimidation also did not exhibit associations with emotional injury.

Actions against online harassment change depending on the type of victimization.

Victims used anti-harassment functions provided by platforms (e.g., blocking/reporting accounts and closing comment forms) in public and private spaces.
By contrast, intimidation, dating requests, and sexual harassment were insignificant in triggering the use of anti-harassment functions.
This may be because anti-harassment functions do not remove the fear of intimidation; victims hesitate to block unwanted messages without explicit hostility, such as dating requests and sexual harassment.

Victims of harassment in public spaces and intimidation tended to raise this with others.
They voice this harassment because anyone can observe harassment in public spaces, and intimidation is a crime.
By contrast, harassment in private spaces and sexual harassment had insignificant effect on voicing about victimization, even if these activities were harmful.

\begin{table*}[t!]		
\tabcolsep = 4pt						
\renewcommand{\arraystretch}{0.932}
\centering			
\small
% \hspace{-6.6cm}
  \begin{tabular}{l|rl|rl|rl|rl|rl|rl}
    \toprule
& \multicolumn{2}{|c|}{Platform}&	\multicolumn{2}{|c|}{Family etc.}	&		\multicolumn{2}{|c|}{Talent agency}	&	\multicolumn{2}{|c|}{Counselor}	&	\multicolumn{2}{|c|}{Legal consultant}	&	\multicolumn{2}{|c}{Police}	\\ 

Explanatory	variable	&	Coef.	&	S.E.			&	Coef.	&	S.E.		&	Coef.		&	S.E.			&	Coef.	&	S.E.			&	Coef.	&	S.E.			&	Coef.	&	S.E. \\ \midrule
Intercept	&	-2.08	&	0.25	***	&	-1.95	&	0.53	***	&	-3.67	&	0.85	***	&	-4.28	&	0.64	***	&	-3.40	&	0.47	***	&	-3.73	&	0.52	***\\					
Harassment in public spaces	&	0.93	&	0.22	***	&	0.49	&	0.20	*	&	0.83	&	0.23	**	&	0.86	&	0.37	*	&	1.10	&	0.27	***	&	1.10	&	0.30	***\\					
Harassment in private spaces	&	0.39	&	0.19	*	&		&			&		&			&		&			&		&			&	-0.78	&	0.48	.\\					
Intimidation	&		&			&	0.50	&	0.26	.	&	0.35	&	0.20	.	&	0.42	&	0.22		&	0.46	&	0.19	*	&	0.62	&	0.28	*\\					
Dating request	&		&			&	0.49	&	0.24	*	&		&			&		&			&		&			&		&	\\						
Sexual harassment	&		&			&		&			&		&			&		&			&		&			&		&	\\						
Age	&		&			&		&			&	0.41	&			&		&			&		&			&		&	\\						
Gender: Female	&		&			&	1.95	&	0.58	**	&	1.16	&	0.87		&		&			&		&			&		&	\\		\midrule	
McFadden's $R^2$	&	0.20	&			&	0.16	&			&	0.23	&			&	0.28		&		&	0.36		&		&	0.31		&	\\	\bottomrule				
% Explanatory	variable	&	Coef.	&	S.E.			&	Coef.	&	S.E.		&	Coef.		&	S.E.			&	Coef.	&	S.E.			&	Coef.	&	S.E.			&	Coef.	&	S.E.		\\	\midrule
% Intercept	&	-3.67	&	0.62	***	&	-1.95	&	0.53	***	&	-0.31	&	0.87	***	&	-4.28	&	0.64	***	&	-3.40	&	0.47	***	&	-3.73	&	0.52	***\\								
% Harassment in public spaces	&	1.12	&	0.24	***	&	0.49	&	0.20	*	&	0.72	&	0.23	**	&	0.86	&	0.37	*	&	1.10	&	0.27	***	&	1.10	&	0.30	***\\								
% Harassment in private spaces	&	0.35	&	0.20	.	&		&			&		&			&		&			&		&			&	-0.78	&	0.48	.\\								
% Intimidation	&		&			&	0.50	&	0.26	.	&	0.41	&	0.20	*	&	0.42	&	0.22		&	0.46	&	0.19	*	&	0.62	&	0.28	*\\								
% Dating request	&		&			&	0.49	&	0.24	*	&		&			&		&			&		&			&		&	\\									
% Sexual harassment	&		&			&		&			&		&			&		&			&		&			&		&	\\									
% Age	&		&			&		&			&		&			&		&			&		&			&		&	\\									
% Gender: Female	&		&			&	1.95	&	0.58	**	&	1.50	&	0.92		&		&			&		&			&		&	\\									
% Type: Influencer	&	2.15	&	0.66	***	&		&			&	-2.03	&	0.66	**	&		&			&		&			&		&	\\	\midrule								
% McFadden's $R^2$	&	0.27	&			&	0.16	&			&	0.30	&			&	0.28		&		&	0.36		&		&	0.31		&	\\	\bottomrule																				
\end{tabular}
\caption{Relationships between talking about victimization and its partners.
$^{***}$, $^{**}$, $^{*}$, and $^.$ indicate significant differences at $p\mathrm{-values} \leq 0.001, 0.01, 0.05, \mathrm{and}~0.1$.
Blank cells represent variables not selected by AIC.
}
  \label{tab_talking}
  
\end{table*}																						

To analyze the relationships between voicing victimization and whom they talk to, logistic regressions were conducted, in which the explained variables were whom they talked to; the explanatory variables were types of online harassment; and the control variables were age levels, and gender.
Table~\ref{tab_talking} shows the results of these regressions.
The explanatory and control variables were selected by Akaike's information criterion (AIC). 

Victims talked about harassment in public spaces, regardless of whom they talked to.
By contrast, in the case of talking about harassment in private spaces, they talked to were limited (platforms).
Remarkably, victims did not seem to speak out about this harassment to the police.
Regarding intimidation, victims contacted authorities with concrete legal actions (talent agencies, legal consultants, and police).
Victims also seemed to talk about it to close people (family, friends, and business friends).
Dating request victimization tended to be discussed with close people.
However, no tendencies regarding sexual harassment revealed with whom it was discussed, i.e., it was not selected by AIC.

\section{Discussion}

\subsection{Picture of Online Harassment Victimization}

Japanese famous people were harmed by various forms of online harassment, 
Victimization ratios of direct harassment in public and private places were 43\% and 27\%, respectively.
That of indirect harassment was 49\%.
These victimization ratios are higher than Japanese journalists (21.5\%)\cite{Yamaguchi2023a} and ordinary people (4.7\%)\cite{Yamaguchi2023}.
This suggests a famous risk in Japan even with considering participation bias of our research.
Online harassment victims tended to be harmed by offline stalking, as in the case of certain New Zealand politicians~\cite{Every-Palmer2015}.
% For influencers, the triggers of offline victimization appear to be their online activities on the Internet.

Victims were harmed by online harassment, particularly harassment in public/private spaces and sexual harassment.
Some of the victims were severely mentally stressed and had negative emotions, such as suicidal thoughts, self-harm/wrecking, and substance abuse.
However, the many of them did not stop their weblog/social media activities because it can be important for business~\cite{Bulck2014}.
Thus, victims had to use online media even if they faced online harassment.

Frequent appearances in mass media and having more than a certain number of followers seemed to provide opportunities for offenders to target and harass victims.
By contrast, having many followers seemed to inhibit being targeted for harassment because offenders seem to fear counterattacks from the victims' followers; for example, if victims criticize offenders' accounts and posts, victims' followers may also criticize offenders.
Gender differences in online harassment victimization were not revealed.
This contrasted with offline harassment, in which women were more targeted.
This may be because offline offenders fear physical counterattacks; however, online offenders can harass victims without fear of physical harm~\cite{Chapman1995}.
Thus, offenders may choose targets, without the fear of counterattacks.
This is consistent with the fact that journalists who do not have the protection of an organization are more likely to be targeted~\cite{Yamaguchi2023a}.
Note that the gender difference might have appeared because female participants were more than male participants, i.e., there may be participation bias.

Victims' tendencies of discussing victimization and using anti-harassment functions provided by weblogs and social media were investigated.
A previous study in Japan~\cite{Yamaguchi2023a} pointed out that the number of people who take concrete action when victimized is limited.
We suggest why it is challenging to take concrete actions against online harassment as follows.
Victims appeared hesitant to speak out about harassment through private methods, such as direct messages, compared with comments in public spaces because victims must disclose the abuse against them.
Victims were not required to complain about comments in public spaces because they were inherently observable.
Victims are hesitant to disclose dating requests and sexual harassment because such sexual victimization aggravates victims' stigma anxieties and their fear of public rejection~\cite{Mesch2006,Andalibi2016,Foster2018}.

In the following, we discuss the consistencies and differences with previous studies. 
\cite{Chen2018,Yamaguchi2023a} indicated the importance of potential victims having countermeasures, e.g., training to handle harassment and the organizations protecting them from abuse. In addition to them, our findings suggest that noticing such countermeasures may inhibit offensive behavior. Although studies of famous female victims\cite{Chen2018,Tonami2022} show that sexism is one of the major factors in online harassment, there is no significant difference in terms of gender in online harassment. This suggests that certain vulnerabilities are gender-specific to females and males.

\subsection{Issues of Online Harassment}

We identified six issues in online harassment.

First, with respect to dating requests, sexual harassment, and intimidation, victims do not tend to use anti-harassment functions provided by platforms, such as blocking/reporting offender accounts and closing comment forms.

Second, victims hesitate to discuss dating requests, sexual harassment, or harassment through private messaging.

Third, offenders harm victims indirectly through activities such as sordid gossip, disclosing and fabricating personal information, and sexualized photoshopped images.
The number of victims of indirect harassment was close to that of direct harassment.
Victims and platforms cannot remove these posts easily
because the removal of such content requires complex procedures.

Fourth, people who seem unwilling to fight back tend to be targeted by offenders.
Offenders use the characteristic features of online communication, which remove the fear of physical harm and online backlash.

Fifth, victims receive stressful messages without the sender's malicious intent, for example, sensitive self-disclosure of suicidal feelings.
Removing such comments and banning such accounts by platforms is difficult because this approach usually does not infringe on the terms of service.

Sixth, online harassment causes victims considerable harm.

\subsection{Approaches against Six Issues}

We can consider several approaches to overcome these concerns.

Facilitating re-consideration of posting comments is effective in discouraging offensive posts~\cite{Katsaros2022}.
This approach has been adopted by many platforms on public comment forums\footnote{For example, \url{https://about.fb.com/news/2019/12/our-progress-on-leading-the-fight-against-online-bullying/}, \url{https://blog.nextdoor.com/2019/09/18/announcing-our-new-feature-to-promote-kindness-in-neighborhoods/}, \url{https://blog.youtube/news-and-events/make-youtube-more-inclusive-platform/}, \url{https://medium.com/jigsaw/helping-authors-understand-toxicity-one-comment-at-a-time-f8b43824cf41}, \url{https://newsroom.pinterest.com/en/creatorcode},
\url{https://newsroom.tiktok.com/en-us/new-tools-to-promote- kindness}, and \url{https://ameblo.jp/staff/entry-12612189833.html}.}.
Our survey revealed that victims receive toxic private messages that harm them.
Victims hesitate to speak about such victimizations.
Thus, adopting the re-consideration function for private message forms is effective.

Changing re-consideration policies to offenders depending on the types of online harassment can reduce some types of online harassment.
For example, the incidences of intimidation, in which victims do not use anti-harassment functions provided by platforms, may decrease by indicating that intimidation is a criminal act.
Flagging that comments and messages may include sexual harassment could discourage sexual harassment~\cite{Mcdonald2014,Blackwell2017}.

A step-by-step guide including the definitions of online harassment, and concrete examples should be issued against online harassment to support victims' actions against online harassment and mitigate online harassment.
Such a guide can facilitate the understanding of complex legal procedures for stopping toxic messages.
This guide should clarify the scope of online harassment by defining it and providing concrete examples.
This clarification can reduce victims' hesitancy to initiate actions against online harassers when they understand that this guide defines the received toxic and sexual messages as harassment~\cite{Blackwell2017}.
The guide can affect offenders' behavior.
Offenders assume that attacking celebrities will not have any repercussions because celebrities are socially invulnerable~\cite{Lawson2017,Ouvrein2018,Lee2020}.
Online harassment may decrease if victims become aware that their behavior constitutes harassment~\cite{Blackwell2017}.
Furthermore, cautioning offenders about the risk of online harassment (e.g., banning and litigation) can discourage toxic behavior.
For example, although the anonymity of the Internet facilitates offensive behavior~\cite{Santana2014,Bliuc2018,plos_takano2021}, this guide informs offenders that they can be exposed through legal procedures.
Thus, vulnerable people are assured of countermeasures against offenders.
As a supplemental approach, online legal consultation, i.e., support from specialists, can be helpful.
Displaying attitudes adopted by platforms to reject online harassment also inhibits toxic posts~\cite{Kang2022}.

The guide should share the definition and adverse effects of online harassment with people~\cite{Blackwell2017}.
This could mitigate the stigma of online harassment~\cite{Mesch2006,Andalibi2016,Foster2018}, victim-blaming~\cite{Scott2019,Hand2021}, and propagating offensive behavior~\cite{Bliuc2018,Yokotani2021}.
Educating people accordingly can ameliorate the adverse effects of sensational coverage~\cite{Kang2022}.

Insurances against online harassment can encourage victims to speak out against it.
The insurances provide financial support and lawyer referral services.
This may discourage toxic behavior because of increased litigation risks.

For sensitive messages, for example, suicidal feelings, platforms can place a report button on their portal.
The platform should introduce inquiry counters to senders of these messages, such as mental health counseling.
This measure could be the simplest initial support for the senders of such messages.
This measure could be expected to decrease the victims' stress because it serves the sender's interests.
Such mental support for senders (offenders) can decrease offensive behavior because they tend to have a psychiatric disorder~\cite{Meloy2016}.

Online counseling is effective for victims' mental health in mitigating the injury of online harassment~\cite{Kraus2010}.

These measures have been partially implemented by a Japanese weblog platform, Ameba~\footnote{\url{https://ameblo.jp/staff/entry-12732411709.html}}.
For example, they defined online harassment and the penalties levied by the platform against such harassment. This also prepared a guide for actions against online harassment, provided/introduced inquiry counters, and facilitated re-consideration in private messages in addition to public comment forms.
Other options are also considered.

\subsection{Limitations and Future Works}

Participants consisting of famous people were recruited by a Japanese Internet company.
The number of participants was not sufficiently large.
More representative sampling would generalize the findings of this study and provide greater insight.

Our findings could be applied to others and not just Japanese famous people.
Examining this applicability should contribute to broader online harassment prevention.

The characteristics of online harassment victimization are common across many cultures and countries~\cite{Chen2018}.
Examining our results in various countries and regions could provide more extensive knowledge.

The effectiveness of the proposed approaches on victims, offenders, and society should be examined.

\section{Summary}

We investigated online harassment victimization among famous people and their responses to such victimization.
Several approaches were proposed for inhibiting harassment and mitigating victims' burdens.
The victimization of famous people can negatively affect their social life.
Online spaces, such as the metaverse and live streaming, have numerous highly detailed interactions, which may increase the risk of online harassment~\cite{Wiederhold2022}.
Our research can contribute to platforms establishing support systems against online harassment from various perspectives, such as supporting victims, inhibiting toxic expressions, and educating people.

\bibliographystyle{plos2015}
\newpage

\clearpage
\setcounter{equation}{0}
\setcounter{section}{0}
\setcounter{figure}{0}
\setcounter{table}{0}

\renewcommand{\figurename}{Fig.}
\renewcommand{\thefigure}{A\arabic{figure}}
\renewcommand{\tablename}{Table.}
\renewcommand{\thetable}{A\arabic{table}}

\renewcommand{\theequation}{S\arabic{equation}}

% \vspace*{\stretch{1}}
% \centering\LARGE{Appendix}
% \vspace{\stretch{2}}
% \pagebreak
% \begin{figure*}[b!]
% \vspace*{\stretch{1}}
% \centering\LARGE{Appendix}\\
% \large{Online Harassment of Japanese Celebrities and Influencers}
% \vspace{\stretch{2}}
% \vspace{10mm}

% \centering
%   \includegraphics[width=1.3\columnwidth]{survey_ss.png}
%    \caption{Screenshot of the survey of this research.
%    See Fig.~\ref{fig_survey_ss_en} for the English version.
% }
%   ~\label{fig_survey_ss}
% \end{figure*}

% \begin{figure*}[t!]
% \centering
%   \includegraphics[width=1.3\columnwidth]{survey_ss_en.png}
%    \caption{English version of the screenshot of the survey of this research.
%    We did not use this version because all survey were conducted in Japanese (Fig.~\ref{fig_survey_ss}).
%    }
%   ~\label{fig_survey_ss_en}
% \end{figure*}

\end{document}